\documentclass[iicol]{sn-jnl}
\usepackage{siunitx}
\usepackage{float}
\usepackage{multicol}
\usepackage{graphicx}%
\usepackage{multirow}%
\usepackage{tabularx}
\usepackage{amsmath,amssymb,amsfonts}%
\usepackage{amsthm}%
\usepackage{mathrsfs}%
\usepackage{esvect}%
\usepackage[title]{appendix}%
\usepackage{xcolor}%
\usepackage{textcomp}%
\usepackage{manyfoot}%
\usepackage{booktabs}%
\usepackage{algorithm}%
\usepackage{algorithmicx}%
\usepackage{algpseudocode}%
\usepackage{listings}%

 \usepackage[numbers,super]{natbib}
\usepackage[T1]{fontenc}
\usepackage[utf8]{inputenc}
\usepackage{hyperref}

\usepackage{soul}

\begin{document}

\title[Article Title]{Magnetic aftereffect and Barkhausen effect in thin films of the altermagnetic candidate Mn$_5$Si$_3$}

\author*[1]{\fnm{Gregor} \sur{Skobjin}}\email{gregor.skobjin@uni-konstanz.de}
\author[2]{\fnm{Javier} \sur{Rial}}
\author[3,4]{\fnm{Sebastian} \sur{Beckert}}
\author[3,5]{\fnm{Helena} \sur{Reichlova}}
\author[2]{\fnm{Vincent} \sur{Baltz}}
\author[6]{\fnm{Lisa} \sur{Michez}}
\author[1]{\fnm{Richard} \sur{Schlitz}}
\author[1]{\fnm{Michaela} \sur{Lammel}}
\author*[1]{\fnm{Sebastian T.B.} \sur{Goennenwein}}\email{sebastian.goennenwein@uni-konstanz.de}

\affil[1]{\orgdiv{{Department of Physics}}, \orgname{University of Konstanz}, \orgaddress{\postcode{{78457}} \city{{Konstanz}},  \country{{Germany}}}}

\affil[2]{ \orgname{Univ. Grenoble Alpes, CNRS, CEA, Gren. INP, IRIG-SPINTEC}, \orgaddress{ \city{{Grenoble}}, \country{{France}}}}

\affil[3]{\orgdiv{{Institute of Solid State and Materials Physics and W\"{u}rzburg-Dresden Cluster of Excellence ct.qmat}}, \orgname{TUD University of Technology Dresden}, \orgaddress{\postcode{{01062}}  \city{{Dresden}}, \country{{Germany}}}}

\affil[4]{\orgdiv{{Dresden Center for Nanoanalysis}}, \orgname{cfaed, TUD University of Technology Dresden}, \orgaddress{\postcode{{01069}}  \city{{Dresden}}, \country{{Germany}}}}

\affil[5]{\orgdiv{{Institute of Physics}}, \orgname{Czech Academy of Sciences}, \orgaddress{\city{{Prague}}, \postcode{{162 00 Praha 6}}, \country{{Czech Republic}}}}

\affil[6]{ \orgname{Aix Marseille Universit\'{e}, CNRS, CINAM, AMUTECH}, \orgaddress{ \city{{Marseille}}, \country{{France}}}}

\abstract{
Altermagnetism as a third distinct type of collinear magnetic ordering lately attracts vivid attention.
We here study the Hall effect response of micron-scale Hall bars patterned into Mn$_5$Si$_3$ thin films, an altermagnet candidate material. Recording transport data as a function of time, at fixed magnetic field magnitude, we observe a time-dependent relaxation of the Hall voltage qualitatively and quantitatively similar to the magnetic viscosity response well established in ferromagnetic films. In addition, the Hall voltage time traces feature clear unilateral steps, which we interpret as Barkhausen steps, i.e., as experimental evidence for abrupt reorientations of magnetic (Hall vector) domains in the altermagnetic candidate material.  
A quantitative analysis yields a Barkhausen length of around 18\,nm in the Hall bar devices with the smallest width of 100\,nm. 
}

\keywords{Altermagnetism, Anomalous Hall Effect, Mn$_5$Si$_3$, Barkhausen Effect\newpage}

\maketitle

The discovery of altermagnetic order\cite{TheOGAltermagnetDude, vsmejkal2020crystal, mazin2021prediction, yuan2020giant, hayami2019momentum} has opened intriguing perspectives, e.g., in the  research fields of spintronics, spin caloritronics, and superconductivity. 
Similar to conventional antiferromagnets, altermagnets feature a compensated spin structure with a vanishing (or very small) net magnetization in real space.\cite{TheOGAltermagnetDude, neel1936proprietes} 
In reciprocal space, however, the band structure of altermagnets shows alternating spin splitting.\cite{TheOGAltermagnetDude,lee2024broken}
Owing to their particular symmetry properties,  altermagnetic materials
can exhibit effects traditionally associated with  ferromagnetic order, including magnetoresistive effects like the giant magnetoresistance and tunneling magnetoresistance,\cite{TheOGAltermagnetDude, chi2024crystal, vsmejkal2022giant, xu2023spin} the occurrence of spin-polarized currents,\cite{TheOGAltermagnetDude} the anomalous Nernst effect,\cite{TheOGAltermagnetDude} and the anomalous Hall effect (AHE).\cite{TheOGAltermagnetDude, RevModPhys.82.1539}

\begin{figure}[h]
    \centering
\includegraphics[width=0.45 \textwidth]{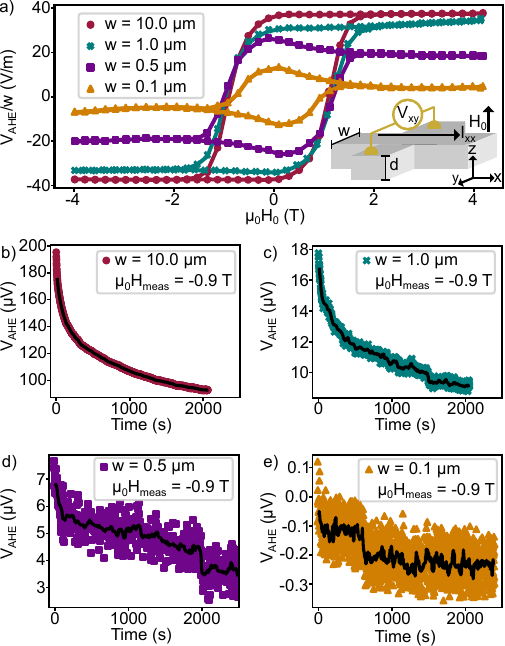} 
\caption{a) Hall voltage hysteresis recorded in $\mathrm{Mn}_5\mathrm{Si}_3$ thin film Hall bars with a width $w =$ 10\,$\mu$m (red), 1\,$\mu$m (teal), 500\,nm (violet) and 100\,nm width (yellow) at a sample temperature of $130\,\mathrm{K}$. A linear ordinary Hall component and a constant offset has been subtracted from all curves. b)-e) Anomalous Hall voltage measurements as a function of time at 130\,K and $-0.9\,\mathrm{T}$ after prior saturation at 4\,T, for the same Hall bars, shown in panel a). Applying a rolling median smoothes the data and brings out Barkhausen steps more clearly (black lines).}\label{Hyst_Susz}
\end{figure}

The Hall voltage response of altermagnetic conductors $V_{xy}=V_{\mathrm{OHE}}+V_{\mathrm{AHE}}$ is usually written as a sum of an ordinary Hall effect (OHE) response scaling with the externally applied magnetic field $H_0$ taken to be along the $z$-direction, and an anomalous Hall effect (AHE) response scaling with the so-called (altermagnetic) Hall vector $h_{\mathrm{Hall}}$\cite{vsmejkal2020crystal, RevModPhys.82.1539, leiviska2024anisotropy}
\begin{equation}
V_{xy} = j_{x} w \mu_{0}(R_{\mathrm{O}} H_{0} + R_{\mathrm{A}} h_{\mathrm{Hall}}). \label{eqAHE}
\end{equation}
Here, $R_{\mathrm{O}}$ and $R_{\mathrm{A}}$ are the ordinary and anomalous Hall coefficients, respectively, $j_{x}$ is the current density applied, $w$ is the width of the conductor across which the Hall voltage $V_{xy}$ transverse to the current flow 
is recorded, and $\mu_0$ is the vacuum permeability constant. 
Maximum Hall voltage arises when $\mathbf{H_0}$ or $\mathbf{h_{Hall}}$, respectively, are orthogonal to both current flow and voltage probing directions, i.e., along the the $z$-direction in the notation used in Fig.\,\ref{Hyst_Susz}a). 

Note also that the Hall voltage response of a ferromagnet with finite magnetization $M$ is captured by Eq.\,(\ref{eqAHE}) upon taking $\mathbf{h_{Hall}} = \mathbf{M}$ .\cite{jiang2010scaling}

As evident from Eq.\,(\ref{eqAHE}), Hall voltage measurements allow investigating the properties associated with long-range magnetic order encoded in  $\mathbf{h_{Hall}}$. 
In ferromagnets, the AHE therefore is routinely used for studying the magnetic hysteresis $M(H_0)$ response.\cite{renuka2024rethinking, tang2016anomalous, aronzon2011room}
Additionally, studying the time-dependent evolution of $V_{xy}(t)$ at constant $H_0$ gives access to the magnetic aftereffect \cite{diao2010magnetic, nabben2024magnetization} and the Barkhausen effect.\cite{barkhausen1919zwei, alessandro1990domain}
The former describes a change of the Hall voltage over time at constant field strength, arising in particular in the range of fields for which the magnetic susceptibility is large\cite{beckert2024detecting,xi2008slow}. In the latter, abrupt reorientations of magnetic domains result in step-like changes of the measured Hall voltage.\cite{barkhausen1919zwei, alessandro1990domain} 
Vice versa, the amplitude of a Barkhausen voltage step $\Delta V_{\mathrm{AHE}}$ can be related to the  magnetic volume flipping its magnetization orientation, e.g., due to magnetic domain nucleation and/or domain wall propagation. 
The observation of Barkhausen steps thus is a smoking gun for the existence of magnetic domains. 
To date, however, neither the magnetic aftereffect nor the Barkhausen effect have been studied in altermagnetic materials.

Here, we report the experimental observation of both the magnetic aftereffect and the magnetic Barkhausen effect in the altermagnetic candidate Mn$_5$Si$_3$. We furthermore use the Barkhausen steps observed in the Hall voltage response as a tool to probe and quantify potential altermagnetic domain features. 
In Hall bar devices with various width $w$ micropatterned into M$_5$Si$_3$ films, we find that bars with a large $w$ are best suited for quantifying the magnetic aftereffect, while the  Barkhausen effect is more prominent in devices with small $w$. 
From the amplitude of the Barkhausen voltage steps observed in the devices with the smallest $w$, we infer an upper boundary for the magnetic volume switching in one Barkhausen step. More specifically, we determined this edge length called the Barkhausen length $L_{\mathrm{Bark}}\lesssim 18\,\mathrm{nm}$, assuming a square domain with volume $V_{\mathrm{Bark}} = L_{\mathrm{Bark}}^2 \cdot d$ extending over the whole film thickness.

Crystalline Mn$_5$Si$_3$ thin films with a thickness of $d$\,=\,19.3\,nm were deposited onto Si(111) single crystal substrates via molecular beam epitaxy using the method described by Kounta et al.\cite{kounta2023competitive}
The films were patterned into Hall bars using electron beam lithography and ion beam etching as detailed in Rial et al.\cite{rial2024altermagnetic}
We here consider Hall bars featuring four different widths
$w=\qtylist{0.1; 0.5; 1; 10}{\micro\meter}$, respectively. 
As the transverse voltage leads also have the width $w$, the sample volume probed in our Hall experiments is roughly $Y_{\mathrm{active}}=w^2 d$.
We refer to $Y_{\mathrm{active}}$ as the active volume 
in the following.

In the Hall effect measurements, a DC charge current $I_x$ was applied to the Hall bar devices and the ensuing transverse voltage $V_{xy}$ was recorded, as sketched in the inset of Fig.\,\ref{Hyst_Susz}a). 
Simultaneously, an external magnetic field $H_0$ normal to both current and voltage was applied using a superconducting magnet cryostat.  
The coordinate system was chosen such that the current, Hall voltage and magnetic field are along $x$, $y$, and $z$ respectively. 
We adjusted the magnitude of $I_x$ such that the current density $j_x = 2.5\times10^{-10}\,\mathrm{A/m}^2$ was the same for all devices, independent of $w$. 
Moreover, we exploited the current reversal technique\cite{beckert2024detecting} in order to suppress spurious thermoelectric contributions to the voltage signal.
In other words, for a given magnetic field magnitude,
two consecutive voltage readings were recorded, one
$V_{xy,+}$ for positive and one $V_{xy,-}$ for negative current polarity, respectively. 
The voltage linear in current is then calculated as $V_{\mathrm{\Delta -}}=\frac{1}{2}(V_{xy,+} - V_{xy,-})$.\cite{beckert2024detecting}

To record  Hall voltage hysteresis loops, we swept the external magnetic field $\mathrm{\mu}_0 H_0 $ applied to the sample from $4\,\mathrm{T}$ to $-4\,\mathrm{T}$ and back. In the following, we focus on experiments performed with the sample temperature stabilized at $T=130\,\mathrm{K}$, well within the presumed altermagnetic phase of Mn$_5$Si$_3$.\cite{kounta2023competitive, reichlova2024observation} 
Typical Hall hysteresis loops for devices with different bar widths $w$ are shown in Fig.\,\ref{Hyst_Susz}a). 
Note that for all Hall data shown in this figure, the ordinary Hall effect (linear in magnetic field), as well as an offset (as commonly seen in Hall voltage measurements,\cite{degrave2013general, badura2023even}) have already been subtracted to bring out the AHE response. 
Consequently, we plot $V_{\mathrm{AHE}}(H_0)=V_{\mathrm{\Delta -}}(H_0)- a\cdot \mu_0 H_0 - b$ in Fig.\,\ref{Hyst_Susz}, where $a$ and $b$ are device-specific constants, as compiled in Tab. \ref{Tab:Para}.

\begin{table}
\begin{tabular}{ |p{0.8cm}||p{1.1cm}|p{1.1cm}|p{1.4cm}|p{1.1cm}|}
 \hline

  $w$ & V$_{\mathrm{sat}}/w$ &$a$&$a/w$&$b$\\
  ($\mu$m) & (V/m) & ($\mu$V/T)& ($\mu$V/T$\cdot \mu$m)& (mV) \\
 \hline
 10   &   37.53  & -133.8 & -13.4&   -6.02\\
 1   &  33.98    & -16.3 &  -16.3& -13.6\\
 0.5   & 19.26    & -10.2 &  -20.4& 1.20\\
 0.1   & 5.40   & -2.6 &  -26.0&  0.13\\
 \hline
\end{tabular}
\caption{ Hall transport parameters extracted from Hall bar devices of different widths $w$. $V_{\mathrm{sat}}$ is the anomalous Hall voltage at saturation,  and the device-specific constants $a$ and $b$ represent the ordinary Hall constant and the voltage offset, respectively. Normalization to the Hall bar width $w$ allows for a better comparison of devices with different dimensions.}\label{Tab:Para}
\end{table}

As evident from Fig.\,\ref{Hyst_Susz}, the Mn$_5$Si$_3$ Hall bars all show an AHE hysteresis saturating at magnetic fields $\mu_0 H_0 \gtrsim 1.5\,\mathrm{T}$, as expected from the literature.\cite{leiviska2024anisotropy,kounta2023competitive,reichlova2024observation}
However, the AHE amplitudes decrease with decreasing bar width $w$, which we ascribe to a modification of the material properties upon patterning the films to micron scale dimensions. 
This is corroborated by the $w$-dependent OHE magnitude (see Tab.\,\ref{Tab:Para}). 
Note also that for small $w$, an additional shoulder appears in the AHE hysteresis loop, which is usually interpreted as evidence for a topological Hall effect contribution
.\cite{reichlova2024observation, neubauer2009topological}

We now turn to the magnetic relaxation measurements for different $w$, again all taken at $T$=130\,K. 
For every relaxation measurement, we first applied $\mu_0 H_0=+4\,\mathrm{T}$ to prepare a defined magnetic state, and then swept the magnetic field to a value $H_{\mathrm{meas}}$ within the high susceptibility region. 
Keeping the magnetic field magnitude fixed at $H_{\mathrm{meas}}$, we then recorded $V_{xy}(t)$ as a function of time $t$ with one data point every $1.5\,\mathrm{s}$, for a time period of $30\,\mathrm{min}$.  For a direct comparison between hysteresis and relaxation data, we again subtract the ordinary Hall effect contribution and offset voltage, using the  appropriate constants $a$ and $b$ (cp. Tab.\,\ref{Tab:Para}). We thus consider $V_{\mathrm{AHE}}(t) = V_{\mathrm{\Delta -}}(t)- a\cdot \mu_0 H_{\mathrm{meas}} - b$ in the following.
Specifically, relaxation data were taken in this way for magnetic field magnitudes of $\mu_0 H_{\mathrm{meas}} = -0.8,-0.9,-1.0,-1.1,-1.2\,\mathrm{T}$ in Hall bar structures with different $w$, of which exemplary data for $\mu_0 H_0 = -0.9$\,T are shown in Figure \ref{Hyst_Susz}b) - e)  (circles, crosses, squares and triangles for $w$ = 10, 1, 0.5 and 0.1\,$\mu$m,  respectively).

The voltage response in Hall bars with $w \ge 1\,\mathrm{\mu m}$ (Fig.\,\ref{Hyst_Susz} panels b) and c)) is dominated by relaxation, which is characterized by a substantial and quasi-continuous decay of $V_{\mathrm{AHE}}(t)$ with time. 
In the narrower Hall bars (Fig\ref{Hyst_Susz} panels d) and e)), step-like features become evident. Averaging the Hall voltage data with a rolling median with a point interval of 30 data points (black lines) brings out the steps in the AHE voltage more clearly. 
We attribute these abrupt steps in $V_{\mathrm{AHE}}(t)$ to the magnetic Barkhausen effect, i.e., to the presence of magnetic domains or domain-like effects in the spatial distribution of the Hall vector $\mathbf{h_{Hall}}$ giving rise to the AHE in Mn$_5$Si$_3$.

\begin{figure}[h]
    \centering
    \includegraphics[width=1\linewidth]{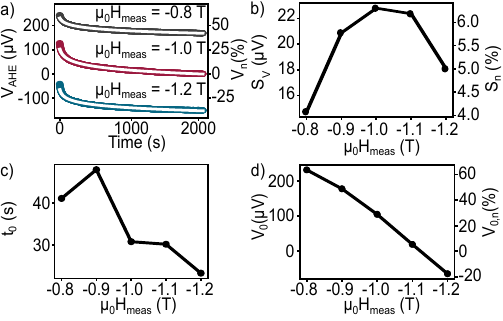}
    \caption{ Typical time-dependent anomalous Hall voltage relaxation data for a Hall bar with $w = 10\,\mathrm{\mu m}$ at $-0.8$\,T (grey), $-1.0$\,T (red) and $-1.2$\,T (blue), together with fits  to Eq.\,\eqref{MagRelaxRes} (white) are given in panel a). The fit parameters $S_V$, $t_0$ and $V_0$ including the normalized $S_{\mathrm{n}}$ and $V_{\mathrm{0,n}}$ extracted from fitting the data from panel a) with Eq.\,\eqref{MagRelaxRes}, are given in panels b) - d). }\label{fig:howtodata}
    
\end{figure}

We first analyze the time-dependent relaxation in the AHE in a Hall bar with $w = 10\,\mathrm{\mu m}$, in which $V_{\mathrm{AHE}}(t)$ is comparably smooth (see Fig.\,\ref{Hyst_Susz}b)).
Figure \ref{fig:howtodata}a) displays relaxation time traces recorded at three different magnetic field magnitudes.
We fit the data with \cite{xi2008slow,jiang2003adjacent, morgunov2017magnetic}
\begin{equation}
V_{\mathrm{AHE}}(t) = V_0 - S_{\mathrm{V}} \, \mathrm{ln}\Bigl(1+\frac{t}{t_\mathrm{0}}\Bigr), \label{eqAHErelax}
\end{equation}
where $V_0=V_{\mathrm{AHE}}(t=0)$, $S_{\mathrm{V}}$ is the so-called magnetic viscosity, and $t_0$ the relaxation time.\cite{beckert2024detecting, xi2008slow, pasko2020temperature}
For comparison between materials, the magnetic relaxation is often normalized to the state of saturation.\cite{beckert2024detecting,lindemuth2004anomalous} We thus normalize 
$V_{\mathrm{AHE}}$(t) to the AHE swing
\begin{equation}
V_{\mathrm{sat}}=\frac{ V_{\mathrm{AHE}}(\mu_0 H_0=+4\,\mathrm{T})-V_{\mathrm{AHE}}(\mu_0 H_0=-4\,\mathrm{T}) }{2}
\end{equation}
and consider 
\begin{equation}
V_{\mathrm{n}}(t) = \frac{V_{\mathrm{AHE}}(t)}{V_{\mathrm{sat}}} = V_{\mathrm{0,n}} - S_{\mathrm{n}} \cdot \mathrm{ln}\Bigl(1+\frac{t}{t_\mathrm{0}}\Bigr). \label{MagRelaxRes}
\end{equation}

Fitting the $V_{\mathrm{AHE}}$(t) data taken at different magnetic fields $\mu_0H_{\mathrm{meas}}$ in Fig.\,\ref{fig:howtodata} a) with Eq.\,\eqref{eqAHErelax} yields the white lines depicted in the same panel. The fits reproduce the data well, as evidenced by consistently high $R^2$-values above 0.98 for a Hall bar width of $w=10\,\mathrm{\mu m}$. A maximum $S_{\mathrm{n}}=6.4\,\%$ is extracted at $\mu_0H_{\mathrm{meas}} = -$1\,T, correlating with the highest susceptibility (largest slope) in the AHE hysteresis in Fig.\,\ref{Hyst_Susz}a) as typical for slow magnetic relaxation.
\cite{lottis1991model}
The characteristic time constant $t_0$ is of the order of several ten seconds, decreasing towards higher field values (Fig.\,\ref{Hyst_Susz}c)). This is to be expected due to higher magnetic fields lowering the energy barrier affecting the magnitude of $t_0$\cite{jiang2003adjacent}.
Both parameters $S_{\mathrm{n}}$ and $t_0$ are comparable in magnitude to those reported for ferromagnetic thin films with perpendicular magnetic anisotropy, e.g., $S_{\mathrm{n}}=2.4\,\%$ and $t_0=0.9\,\mathrm{s}$ in Pt/Co/AlO$_{\mathrm{x}}$, and $S_{\mathrm{n}}=5.5\,\%$ and $t_0=13.5\,\mathrm{s}$ in MnAl, respectively.\cite{beckert2024detecting} In other words, the AHE relaxation in Mn$_5$Si$_3$ is qualitatively and quantitatively similar to the one observed in conventional ferromagnets. While our Mn$_5$Si$_3$ Hall bar devices with
$w < 10\,\mu$m also display similar trends, we here refrain from a quantitative analysis, as the more pronounced Barkhausen steps substantially reduce the fit quality with $R^2$ values decreasing down to 0.5 for $w=0.1\,\mathrm{\mu m}$.

In a conventional ferromagnet, the Barkhausen steps are attributed to abrupt changes of the magnetization orientation within some portion of the magnetically active material which we denote the Barkhausen volume $Y_{\mathrm{Bark}}$.\cite{mustaghfiroh2024temperature, kirilyuk1996barkhausen} 
For the $\mathrm{Mn}_5\mathrm{Si}_3$ film of interest here, we assume a similar connection between the Barkhausen voltage step amplitude $\Delta V_{\mathrm{Bark}}$ and a local reorientation of the Hall vector $\mathbf{h_{Hall}}$. The ratio between 
$\Delta V_{\mathrm{Bark}}$ and the total change in voltage $\Delta V_{\mathrm{AHE}}=2 V_{\mathrm{sat}}$ from saturation to saturation within a full AHE hysteresis then represents a measure for the fraction of the active volume $Y_{\mathrm{active}}$
in which the Hall vector changes sign during the Barkhausen step. For comparison between Hall bars with different widths $w$ and a limited thickness $d$, we express the reorientation of the magnetic material during the Barkhausen step in terms of the  Barkhausen length $L_{\mathrm{Bark}}$, 
\begin{equation}
 \frac{Y_{\mathrm{Bark}}}{ Y_{\mathrm{active}}} = \frac{\Delta V_{\mathrm{Bark}}}{\Delta V_{\mathrm{AHE}}} = \frac{L_{\mathrm{Bark}}^2\cdot d}{w^2 \cdot d}.\label{ConversionVol}
\end{equation}

\begin{figure}[ht]
    \centering
\includegraphics[width=1\linewidth]{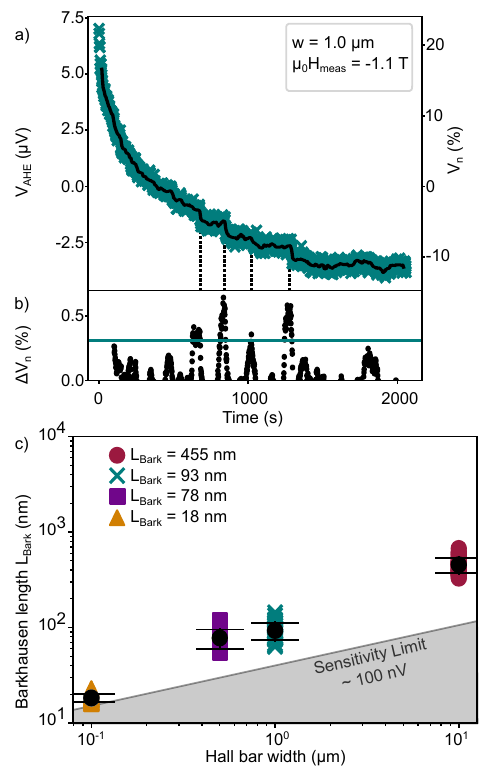} 
\caption{a) Schematic visualization of the data analysis process used to detect and quantify Barkhausen voltage steps $\Delta V_{\mathrm{Bark}}$, using AHE voltage relaxation data from a Hall bar with $w = 1\,\mathrm{\mu m}$ taken at $\mu_0H_{\mathrm{meas}} = -1.1\,\mathrm{T}$. The initial $V_{\mathrm{AHE}}(t)$ relaxation data (teal) are smoothed via a rolling median (black line) and a relaxation fit is subtracted. Numeric differentiation, a rolling sum and a multiplication with the individual time steps between data points yields $\Delta V_{\mathrm{n}}(t)$ data in which Barkhausen steps appear as local maxima. All local maxima in $\Delta V_{\mathrm{n}}(t)$ above a threshold defined by three time the standard deviation of $\Delta V_{\mathrm{n}}(t)$ are taken as Barkhausen steps with amplitude $\Delta V_{\mathrm{Bark}}$. c) The Barkhausen lengths $L_{\mathrm{Bark}}$ extracted from the relaxation time traces appears to scale with the Hall bars widths. The corresponding average Barkhausen lengths $L_{\mathrm{Bark}}$ are quoted in the legend. The grey line shows the experimental sensitivity limit (for details see text).}\label{Golden}
\end{figure}

For a quantitative evaluation of $L_{\mathrm{Bark}}$, we extract the voltage changes $\Delta V_{\mathrm{Bark}}$ corresponding to Barkhausen steps from the $V_{\mathrm{AHE}}(t)$ data as depicted in Fig.\,\ref{Golden}.  
First, a rolling median with an interval of 30 data points is applied to the $V_{\mathrm{AHE}}(t)$ data to smooth the noise 
while preserving steps as typical for the Barkhausen effect. The result is shown as a black line in Fig.\,\ref{Golden}a). 
We then subtract a fit to the data using Eq.\,\eqref{eqAHErelax} to remove the smooth time-dependent changes arising from relaxation. The resulting residual voltage $V_{\mathrm{res}}(t)=V_{\mathrm{AHE}}(t)-V_{\mathrm{fit}}(t)$ is then numerically differentiated as $\frac{\Delta V_{\mathrm{res}}(t)}{\Delta t}$. To account for Barkhausen steps spanning more than one time step (more than two data points), a rolling sum with an 
interval of 30 points is applied to the data. Finally, we multiply with $\Delta t / \Delta V_{\mathrm{AHE}}$ to obtain the normalized $\Delta V_{\mathrm{n}}(t)$  shown in Fig.\,\ref{Golden}b). The time interval associated with the rolling sum and rolling median were chosen to be of the same size and such that only one potential Barkhausen step would typically be contained.
All local maxima in $\Delta V_{\mathrm{n}}(t)$ that exceeded three times the standard deviation of $\Delta V_{\mathrm{n}}(t)$ then were selected as Barkhausen steps $\Delta V_{\mathrm{Bark}}$. 
For the data shown in Fig.\,\ref{Golden}a), this evaluation procedure results in the identification of four Barkhausen steps, as indicated by the vertical dashed lines.  

All Barkhausen step amplitudes $\Delta V_{\mathrm{Bark}}$ extracted from different datasets following the procedure just described were converted into a Barkhausen volume $Y_{\mathrm{Bark}}$ and a Barkhausen length $L_{\mathrm{Bark}}$, respectively, using Eq.\,\eqref{ConversionVol}.
Figure \ref{Golden}c) shows these data. The average Barkhausen length $L_{\mathrm{Bark}}$ decreases 
from $455\,\mathrm{nm}$ for $w=10\,\mathrm{\mu m}$ 
down to $18\,\mathrm{nm}$ for $w=0.1\,\mathrm{\mu m}$.
While the order of magnitude of $L_{\mathrm{Bark}}$ appears reasonable in view of the typical micrometer to nanometer domain sizes reported in both ferromagnetic and antiferromagnetic thin films,\cite{dugato2019magnetic, ando2016modulation, shiratsuchi2018observation, fitzsimmons2008antiferromagnetic, cheong2020seeing} 
the strong dependence of $L_{\mathrm{Bark}}$ on the Hall bar width $w$ is  surprising. Indeed, for a typical Barkhausen process, one would expect an average $L_{\mathrm{Bark}}$ reflecting the domain nucleation and/or growth characteristics governed by the free enthalpy landscapes,\cite{schwarz2004visualization, hubert2008magnetic} such that $L_{\mathrm{Bark}}$ should not significantly depend on the lateral device dimensions. 

On the other hand, micropatterning a sample can influence its domain structure, as reported also in antiferromagnetic thin films, for example due to edge-induced anisotropy influencing the stability of magnetic domains.\cite{reimers2023magnetic,moritz2002domain, amin2024altermagnetism}
As the altermagnetic phase in Mn$_5$Si$_3$ thin films is stabilized by strain,\cite{kounta2023competitive} a possible relaxation of the lattice arising from the micropatterning could result in a change of the (alter)magnetic properties.
This would impact smaller devices more, as the surface to volume ratio changes with device dimension $w$. 
Possibly, such an effect could also explain the change in the AHE hystersis shape with device width $w$ evident in Fig.\,\ref{Hyst_Susz}a).

We rationalize the dependence of $L_{\mathrm{Bark}}(w)$ on device dimensions in Fig.\,\ref{Golden}c) as follows: The magnitude of the AHE voltage $V_{\mathrm{AHE}}(w)\propto w$ scales linearly with the Hall bar width (see Eq.\,\eqref{eqAHE}). 

Inserting $Y_{\mathrm{Bark}}=L_{\mathrm{Bark}}^2 d$ in Eq.\,\eqref{ConversionVol} and substituting $\Delta V_{\mathrm{AHE}}$ with the anomalous Hall contribution of $V_{xy}$ from Eq.\,\eqref{eqAHE} yields $\Delta V_{\mathrm{Bark}}=j_x w R_{\mathrm{A}}  \frac{Y_{\mathrm{Bark}}}{Y_{\mathrm{active}}}$.
Using $Y_{\mathrm{active}}=w^2 d$ as introduced above, this implies that  $Y_{\mathrm{Bark}} =\Delta  V_{\mathrm{Bark}} \frac{ w d}{ R_{\mathrm{A}} j_x }$. 
The resolution limit for $\Delta  V_{\mathrm{Bark}}$ is given by the inevitable noise floor of the electrical detection circuit.
Taking this noise floor to be of the order of $100\,\mathrm{nV}$, such that  only $\Delta V_{\mathrm{Bark}}\gtrsim 100\,\mathrm{nV}$ can be resolved in experiment, one obtains the $L_{\mathrm{Bark}}$ sensitivity limit shown by the grey line in Fig.\,\ref{Golden}c).
Vice versa, this implies that the Hall vector Barkhausen volume $Y_{\mathrm{Bark}}$ characteristic for Barkhausen-type processes in $\mathrm{Mn}_5\mathrm{Si}_3$ films should be extracted from the devices with the smallest $w$, since these are most sensitive to Barkhausen-type effects. 

The Barkhausen voltage steps observed in the Hall bar device with the smallest width $w= 0.1\,\mathrm{\mu m}$ correspond to a concerted AHE-active volume reorientation with a Barkhausen length of $L_{\mathrm{Bark}}=18\,\mathrm{nm}$ 
at maximum, see Fig.\,\ref{Golden}c). 
While this only can be an upper estimate of the typical Hall vector domain size, the dimensions 
are consistent with the structural domains 
reported in previous studies of Mn$_5$Si$_3$ thin films and related phases during thin film growth, ascribed to hexagonal superstructures caused by strain relief of a MnSi phase.\cite{kounta2023competitive, schwinge2005structure, kumar2004thin} 
Possibly, the characteristic $Y_{\mathrm{Bark}}$ magnitude thus is connected to epitaxial strain textures.
\cite{rial2024altermagnetic} Last but not least, the $L_{\mathrm{Bark}}$ extracted here can be compared to the dimensions of the so-called altermagnetic variants reported recently in $\mathrm{Mn}_5\mathrm{Si}_3$ films.\cite{rial2024altermagnetic} In Hall bars with a width $w=0.1\,\mathrm{\mu m}$, only one altermagnetic variant was identified,\cite{rial2024altermagnetic} while we observe multiple Barkhausen steps in such devices. This implies that different types of magnetic (Hall vector) textures must be considered in the discussion of altermagnetic thin films.

In summary, we have observed the magnetic aftereffect and the Barkhausen effect in the altermagnet candidate Mn$_5$Si$_3$. 
Using anomalous Hall effect experiments to study the evolution of the Hall vector in micron-scale Mn$_5$Si$_3$ Hall bars both as a function of magnetic field magnitude and time, we find a relaxation of the Hall voltage with time qualitatively and quantitatively similar to the magnetic viscosity response reported in ferromagnetic films with perpendicular magnetic anisotropy. In addition, the Hall voltage time traces feature abrupt Barkhausen-like steps. We interpret these as experimental evidence for magnetic (Hall vector) domains. 
A quantitative analysis yields a Barkhausen length $L_{\mathrm{Bark}}=18\,\mathrm{nm}$ in the Hall bar devices with the smallest width $w=100\,\mathrm{nm}$. 
Our  findings thus show that domain-like relaxation phenomena also can be at play in altermagnetic materials.

\backmatter

\bmhead{Acknowledgements}

This work was funded by the Deutsche Forschungsgemeinschaft (DFG, German Research Foundation) via projects ID 445976410 and ID 490730630 and via the SFB 1432, project ID 425217212. The materials growth was supported by the French national research agency (ANR) and the Deutsche Forschungsgemeinschaft (DFG), Germany (Project HEXAS - Grant No. ANR-24-CE92-0038-02) and by a French government grant managed by the Agence Nationale de la Recherche, France under the France 2030 program (SPINMAT, ANR-22-EXSP-0007).

\bibliographystyle{sn-nature}  
\bibliography{sn-bibliography}  

\end{document}